\documentclass[aip,pof,reprint]{revtex4-1}

\usepackage{graphicx}
\usepackage{amsmath}
\usepackage{xcolor}
\usepackage[normalem]{ulem}

\newcommand{\be}{\begin{equation}}
\newcommand{\ee}{\end{equation}}
 
\newcommand{\rev}[1]{{#1}} 
\newcommand{\dl}[1]{} 
\newcommand{\revs}[1]{{#1}} 
\newcommand{\dls}[1]{} 

\newcommand{\rme}{\mathrm{e}}
\newcommand{\rmi}{\mathrm{i}}

\newcommand{\bk}{\mathbf{k}}

\begin{document}

\title{Surface waves on arbitrary vertically-sheared currents}
\author{Benjamin K. Smeltzer}
\affiliation{Department of Energy and Process Engineering, Norwegian University of Science and Technology, N-7491 Trondheim, Norway}
\author{Simen \AA. Ellingsen}
\affiliation{Department of Energy and Process Engineering, Norwegian University of Science and Technology, N-7491 Trondheim, Norway}
\date{\today}

\begin{abstract}
We study dispersion properties of linear surface gravity waves propagating in an arbitrary direction atop a current profile of\dl{arbitrary} depth-varying magnitude using a piecewise linear approximation, and develop a robust numerical framework for practical calculation. The method has been much used in the past \rev{for the case of waves propagating along the same axis as the background current}\dl{in 2D}, and we herein extend and apply it to \rev{problems with an arbitrary angle between the wave propagation and current directions}\dl{3D problems}. Being valid for all wavelengths without loss of accuracy, the scheme is particularly well suited to solve problems involving \rev{a broad range of wave vectors, such as ship waves and Cauchy-Poisson initial value problems for example.}\dl{Fourier transformations in the horizontal plane.} We examine the group and phase velocities over different wavelength regimes and current profiles, highlighting characteristics due to the depth-variable vorticity. We show an example application to ship waves on an arbitrary current profile, and demonstrate qualitative differences in the wake patterns between concave down and concave up profiles when compared to a constant shear profile with equal depth-averaged vorticity. 
\dls{New insight is given concerning}\revs{We also discuss} the nature of \dls{extra spurious} \revs{additional} solutions to the dispersion relation when using the piecewise-linear model\dls{, where an analytical solution to a two-layer model is used to explain the solutions as vortical structures drifting downstream at a velocity near that of the artificial interfaces between layers}.  \revs{These are vorticity waves, drifting vortical structures which are artifacts of the piecewise model. They are absent for a smooth profile and are spurious in the present context.}
\end{abstract}

\maketitle

\section{Introduction}
A complete understanding of surface water wave propagation on a background current profile is of great importance in areas within oceanography, marine and coastal engineering, and naval architecture\cite{Pere}. The presence of an underlying current modifies the wave dispersion potentially affecting key quantities such as wave loads on structures, wave propagation near coastlines, or ship wave resistance. Furthermore, measurements of wave frequencies at known wavelengths (e.g. using high-frequency radar) can be used to infer the underlying current profile\cite{StewartJoy,Graber96,Fernandez96,Lund15}, 
relevant for predicting storm surges and understanding the mechanisms of climate change\cite{Lund15}.
Many studies and models in these areas have used simple velocity profiles such as depth-uniform or linear depth dependence, largely due to mathematical tractability as analytical solutions exist only for a select few of these current profiles\cite{Pere}. Various approximation techniques have been developed for more realistic profiles\cite{StewartJoy,KirbyChen,Skop87,SwanJames,Shrira93}, yet these have limited range of applicability. The goal of this work is to demonstrate an approximation method for calculating the dispersion relation on an arbitrary current profile in three dimensions valid for all wavelengths that is suitable for practical calculations by engineers.

For the purposes of this work we consider \dl{small}\rev{infinitesimal}-amplitude surface waves propagating on a background rotational current flow that is \rev{steady,} incompressible\rev{,} and inviscid. \revs{Although} viscosity \revs{is neglected for the wave motion}, viscous effects are certainly involved in \revs{generating the shear current itself. We are not, however, concerned with how the background current may have come about. Thus, in}\dl{In} this small wave amplitude regime, we assume the wave-current interaction to be unidirectional\dls{ in a sense}: the current\dls{s} affect\revs{s} the wave motion but not \dls{visa versa}\revs{\emph{vice versa}}. The current profiles we consider are of depth-variable magnitude yet constant direction \rev{and are assumed to be homogeneous in the horizontal directions}. 
%

The vast majority of the body of work on surface waves and shear currents \rev{considers wave propagation parallel or anti-parallel to the current, which we refer to as two dimensional (2D) with as single vertical and horizontal spatial axis. The generalized case consists of a horizontal plane with waves propagating at an arbitrary oblique angle to the direction of the current, referred to three dimensional (3D).}\dl{is in two dimensions.} \dls{Ellingsen\cite{Ell14,Ellcp} recently showed} \revs{It was recently shown} how 3D solutions to the Euler equations 
in the presence of a linear shear current (constant vorticity)
can be used to solve classical problems such as ship waves and ring waves\revs{\cite{Ell14,Li16,Ellcp}}. 
Shear currents were found to have the potential to significantly alter 
the characteristics of wave propagation in inherently 3D problems, evidenced by the behavior of ship waves as well as solutions to Cauchy-Poisson initial-value problems\cite{Ellcp}. In the former case it was shown that the Kelvin angle (the maximum wake angle with appreciable wave energy) is a function of shear strength and orientation angle of the current relative to ship motion. For initial-value problems it was shown that the difference between phase velocity and group velocity can be very different in propagation directions where waves are assisted or inhibited by the sub-surface shear, 
respectively,
leading to anisotropic behavior in the time evolution of an initial surface disturbance.

For most realistic current profiles however, the vorticity is not constant with depth. To treat profiles with arbitrary current depth-dependence, various approximation techniques have been 
developed, typically involving expansions in a small parameter representing the magnitude of the current velocity relative to the phase velocity of the waves\cite{StewartJoy,KirbyChen,Skop87,SwanJames}, or the departure from a velocity potential solution\cite{Shrira93}. These methods have been used for many practical calculations such as inferring the background current from phase velocity measurements\cite{StewartJoy,Graber96,Fernandez96,Lund15}, yet complications 
occur when applying them to problems involving the entire wave-spectrum as their accuracy 
\dls{suffers} \revs{is difficult to predict \emph{a priori} and can suffer} in certain wavelength regimes\dl{ where the expansion parameter is no longer small}. Many problems such as the above-mentioned ship waves and ring waves are conveniently solved in Fourier space, 
whereupon integration over all horizontal wave vectors is performed, and a fast dispersion calculation method giving the same approximation accuracy over the entire wave-spectrum at \dls{no}\revs{little} extra cost is thus desired.

In this work we use a method based on a piecewise linear approximation (PLA) to the 
background current's velocity profile. The profile is divided into vertical layers each assumed to have constant vorticity. Within each layer, solutions to the linearized Euler equations are found, and these solutions are matched appropriately at the layer interfaces to yield the full solution over the entire domain\cite{DrazinReed}. This method has been extensively used in the past perhaps first by Lord Rayleigh\dls{ in 1892}
\cite{rayleigh1879,rayleigh1892}, mostly with 2-3 layers. The simplest two-layer case with constant shear in the lower layer and constant current in the upper layer (no shear) was analyzed by Thompson\cite{thomp49}, and constant shear in an upper layer on an infinite lower layer of zero current by Taylor\cite{taylor} to investigate the potential of a current produced by 
a bubble curtain used
as a breakwater to stop waves. The generalized two-layer result was later given by Dalrymple\cite{DalrTR}. 
Zhang\cite{Zhang05} compared the PLA method to other approximation methods\cite{StewartJoy,KirbyChen,Shrira01}, showing how it is able to accurately calculate dispersion properties at all wavelengths. \dl{To wit, with}\rev{With} the implementation described herein we calculate phase velocities at the $1\%$ accuracy level or better with just $4$-$5$ layers in the entire wave vector plane, making the method calculationally cheap, conceptually simple, easily implement and hence ideal for Fourier transformation purposes.

The primary difficulty in using the piecewise linear approximation involves extra \dls{unphysical} solutions to the polynomial equations that are solved \revs{in order} to find the phase velocity for a given wavevector\cite{thomp49,Shrira01,Zhang05}. These \dls{spurious} solutions \revs{are spurious in the present context, and} have phase velocities near the velocity of the background flow at the layer interfaces\cite{Zhang05}\dls{, and are due to}\revs{. They are artifacts introduced by} the discontinuities in the shear, something we discuss \dls{in detail} \revs{further} in Section \ref{sec:extrasol}. Despite the complication of discarding the spurious solutions, the piecewise linear approximation has been much used in studying the stability of small disturbances on shear flows using a perturbation type approach\cite{LH98,Cap91,Cap92}. In some cases there is disagreement with work considering similar smooth flows, raising questions about the accuracy of the method\cite{Eng00,Mor98,Shrira01}. Zhang\cite{Zhang05} addressed these issues showing the convergence of the piecewise profile to be $O(\Delta z^2)$ where $\Delta z$ is the layer thickness, and studied the \dls{spurious} \revs{extra} solutions.

As many applications are inherently 3D (initial value problems, ship waves, radiation, refraction) \revs{we} implement and demonstrate the PLA method in 3D. Further analysis of the nature of the \dls{spurious} \revs{extra} solutions is given by considering a simplified two-layer fluid. We demonstrate the convergence and approximation accuracy over a range of wavelength scales as a function of the number of layers. The PLA is then further applied to \dls{calculate} \revs{calculating} the directional dependence of the group and phase velocities on two profiles with non-constant vorticity. Finally, we \dls{show an example application to solving the classical ship wave problem on an arbitrary current profile.} \revs{solve the classical ship wave problem on an arbitrary shear profile as an example application.}

\section{Formulation of the model}
\label{sec:der}

\begin{figure}[ht]
  \includegraphics[scale=1.0]{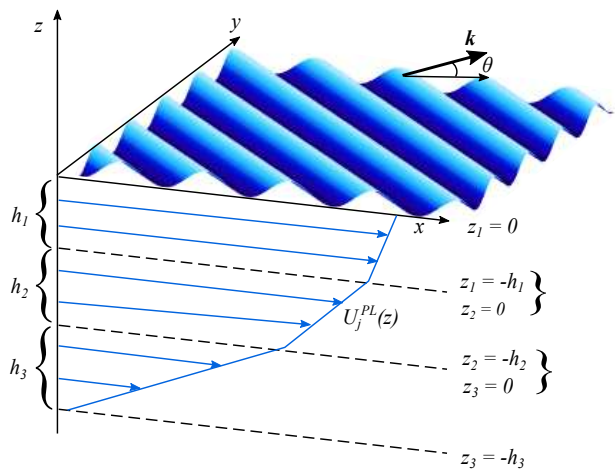}
  \caption{The geometry and definitions used in the text demonstrated for a 3-layer piecewise linear profile. Vertical coordinates $z_j$ are defined within each layer. The current is assumed to be oriented along the $x$-direction, and varies linearly within each layer described by the function $U_j^{PL}(z_j)$.}
\label{fig:geom}
\end{figure}

We consider 3D surface waves propagating on a depth-varying current $U(z)$ oriented along the horizontal $x$-axis. 
The velocity field can be written:
\begin{equation}
\mathbf{v} = \left(U(z) + \hat{u}, \hat{v}, \hat{w}\right)
\end{equation}
following the notation of Ellingsen\cite{Ell14} where hatted quantities are assumed to be small perturbations due to the waves. We assume a progressive surface wave with infinitesimal surface height $\hat{\zeta}(x,y)$ in the horizontal plane, wavevector $\mathbf{k} = (k_x,k_y)$ making an angle $\theta$ to the 
$x$-axis\revs{, pressure $P = -\rho g z + \hat{p}$,}
and frequency $\omega$ such that the velocity and pressure perturbations are \dls{assumed to have the spatial and time dependence} expressed as:
\begin{equation}
\left(\hat{\zeta},\hat{u}, \hat{v}, \hat{w}, \hat{p}\right) = \left(\zeta,u(z),v(z),w(z),p(z)\right)e^{i\left(\mathbf{k}\cdot\mathbf{r}-\omega t\right)},
\end{equation}
where $\mathbf{r}$ is the position vector in the $xy$-plane. We will artificially divide up the fluid column into $N$ layers in the vertical direction so that the vorticity of the background flow be constant inside each layer, as shown in Fig. \ref{fig:geom}. Each layer has a thickness $h_j$ and vertical coordinate within each layer $z_j = z + \sum_{l=1}^{j-1}h_l$ as shown in Fig. \ref{fig:geom}. The approximate current profile in layer $j$ is
\begin{equation}
U_j^{PL}(z_j) = U_{j-1} + S_jz_j,
\end{equation}
where $U_j \equiv U (z = -\sum_{l=1}^jh_l)$ is the value of the non-linear current profile at the layer interfaces, and $S_j = (U_j-U_{j+1})/h_j$ is the layer mean vorticity. 

The linearized Euler and continuity equations inside layer $j$ are
\begin{subequations}
\begin{align}
\left[-i\omega + ik_xU_j^{PL}(z_j)\right]u_j + S_jw_j =& -ik_xp_j/\rho;\label{eq:eulerx} \\
\left[-i\omega + ik_xU_j^{PL}(z_j)\right]v_j =& -ik_yp_j/\rho; \label{eq:eulery}\\
\left[-i\omega + ik_xU_j^{PL}(z_j)\right]w_j =& -\dl{i}p_j^\prime/\rho; \label{eq:eulerz}\\
ik_xu + ik_yv + w' =&0,
\label{eq:cont}
\end{align}
\end{subequations}
where the prime denotes $\partial/\partial z$. In the interior of each layer this system of equations leads to the Rayleigh equation for the vertical velocity component $w_j$
\begin{equation}
w_j^{\prime\prime} - k^2w_j = 0,
\label{eq:rayleigh}
\end{equation}
with general solution
\begin{equation}
w_j = A_j\sinh{k(z_j+h_j)} + B_j\cosh{k(z_j+h_j)},
\label{eq:wj}
\end{equation}
where $k = |\mathbf{k}|$. When $A_j$ and $B_j$ are known, the two other velocity components and the pressure can be found by inserting this solution back into 
Eqs.~\eqref{eq:eulerx}-\eqref{eq:cont}. 
There are thus $2N$ unknowns for $N$ layers. Four 
types of boundary
conditions \revs{for $w$ and $p$} provide the necessary $2N$ equations; \revs{one }
at the bottom\dls{ and}\revs{, one at the} free surface, \dls{as well as} \revs{and $2N-2$} matching conditions \dls{for the velocity and pressure} at the layer interfaces. Considering finite, uniform depth, the vertical velocity component must vanish at the bottom ($z_N = -h_N$),
\begin{equation}
  B_N = 0.
\label{eq:bottom}
\end{equation}
The vertical velocity component $w(z)$ must be continuous everywhere leading to a kinematic boundary condition at the layer interfaces $z_j = h_j$: $w_j(-h_j) = w_{j+1}(0)$, which gives
\begin{equation}
  B_j = A_{j+1}\sinh kh_{j+1} + B_{j+1}\cosh kh_{j+1}.
  \qquad j\in (1,N-1).
  \label{eq:vmatch}
\end{equation}
The second matching condition at the interfaces is the continuity of pressure (a dynamic boundary condition)\rev{. The pressure in each layer}\dl{which} can be formulated in terms of $w_j$ and its derivative
as
\begin{equation}
-k^2\frac{p_j}{\rho} = -ik_xS_jw_j + \left[-i\omega + ik_xU_j^{PL}(z_j)\right]w'_j.
\label{eq:pressure}
\end{equation}
Inserting Eq.~\eqref{eq:wj} yields
\begin{align}
  -k^2\frac{p_j}{\rho} &= \left[-ik_xS_{j}A_{j} + \left(-i\omega + ik_xU_j^{PL}(z_j)\right)kB_{j}\right]\sinh kh_{j} \nonumber\\
  &+\left[-ik_xS_{j}B_{j} + \left(-i\omega + ik_xU_j^{PL}(z_j)\right)kA_{j}\right]\cosh kh_{j}.
\end{align}

Continuity of pressure requires that $p_j(z_j=-h_j) = p_{j+1}(z_{j+1}=0)$. We further insert 
Eq.~\eqref{eq:vmatch} to eliminate coefficients $A_{j\neq 1}$ and express a combined kinematic and dynamic condition at the layer interfaces
in the form
\begin{subequations}
\begin{align}
  k\sigma_1 A_1 + \left[\gamma_1 - \gamma_2 - k\sigma_1\coth kh_2\right]B_1 + k\sigma_1\left[\cosh kh_2\coth kh_2 - \sinh kh_2\right]B_2   &=0 \label{eq:pmatcha}\\
  \left(k\sigma_j/\sinh kh_j\right) B_{j-1} + \left[\gamma_j - \gamma_{j+1} - k\sigma_j\left(\coth kh_j+\coth kh_{j+1}\right)\right]B_j  & \nonumber\\ 
  +k\sigma_j\left[\cosh kh_{j+1}\coth kh_{j+1} - \sinh kh_{j+1}\right]B_{j+1} &= 0,
  \label{eq:pmatchb}
\end{align}
\end{subequations}
for $j\in (2,N-1)$,
where $\gamma_j \equiv -ik_xS_j$ and $\sigma_j  \equiv -i\omega + ik_xU_j$. The final condition is at the free surface $\zeta$, a combined kinematic and dynamic boundary condition, expressed here neglecting surface tension: 
\begin{equation}
A_1\left[\gamma_1\sigma_0\tanh kh_1 + \sigma_0^2k + gk^2\tanh kh_1\right] + B_1\left[\gamma_1\sigma_0 + \sigma_0^2k\tanh kh_1 + gk^2\right] = 0.
\label{eq:dkc}
\end{equation}

Eqs.~(\ref{eq:bottom}), (\ref{eq:pmatcha}), (\ref{eq:pmatchb}), and (\ref{eq:dkc}) for $A_1$ and $B_j$ form an $(N+1)\times (N+1)$ homogeneous linear system with coefficient matrix $\mathbf{M}$. The determinant of $\mathbf{M}$ must be zero for non-trivial solutions of the vertical velocity coefficients to exist, leading to a polynomial equation for the unknown $\omega$. This equation is degree $N+1$, giving in general $N+1$ solutions for the dispersion relation $\omega(\mathbf{k})$, or the phase velocity $C(\mathbf{k}) = \omega \mathbf{k}/k^2$. \rev{A computationally efficient method for finding $\omega(\mathbf{k})$ scalable to many layers ($N>100$) involves formulating the linear system as a quadratic eigenvalue problem, expressing the coefficient matrix $\mathbf{M} = \mathbf{L_0} + \mathbf{L_1}\omega + \mathbf{L_2}\omega^2$. The eigenvalues $\omega$ and corresponding eigenvectors $\mathbf{x}$ of the resulting equation $\left(\mathbf{L_0} + \mathbf{L_1}\omega + \mathbf{L_2}\omega^2\right)\mathbf{x}=0$ can be found using a standard polynomial eigenvalue solver.}


\begin{figure*}[ht]  
  \includegraphics{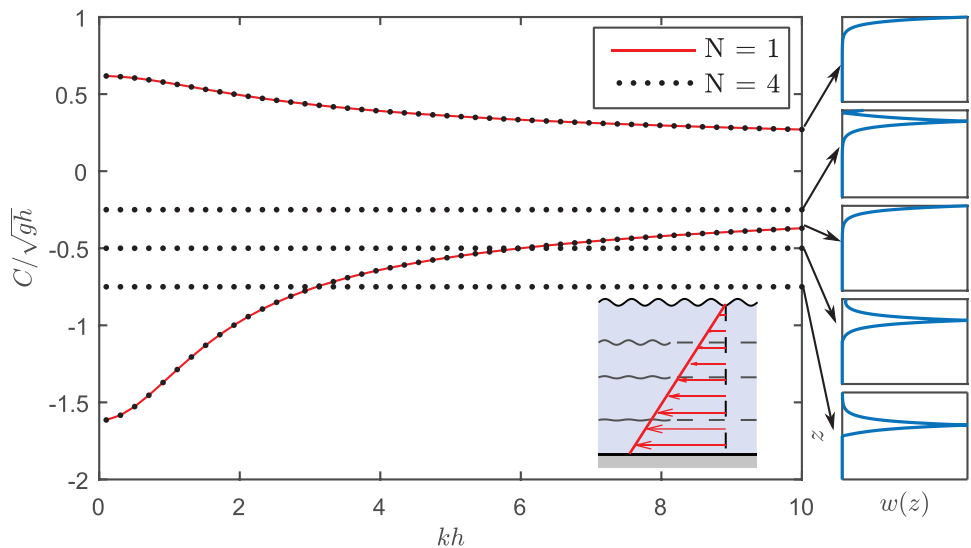}
  \caption{The phase velocity solutions from the piecewise linear approximation as a function of $kh$ ($\theta = 0$) for a linear shear profile $U(z) = z\sqrt{g/h}$ divided artificially into 4 layers \rev{ ($N = 4$) of equal thickness} (inset). \rev{The black dotted lines show the $N+1$ phase velocity solutions of the piecewise linear approximation for a given value of $kh$. Three solutions have phase velocities approximately equal to that of the current profile at the layer interfaces ($U(-h/4)$, $U(-h/2)$, and $U(-3h/4)$).} The analytical result \rev{readily derived from the $N=1$ case} is shown as a solid \rev{red} line to illustrate the known physical solutions. \rev{The small sub-figures show the vertical velocity profiles $w(z)$ corresponding to each of the 5 solutions to the $N=4$ PLA dispersion relation for $kh=10$.}}
\label{fig:roots}
\end{figure*}

\rev{As an illustrative example of the $N+1$ solutions to the PLA dispersion relation, we consider a current profile of constant vorticity $U(z) = z\sqrt{g/h}$ with total depth $h$,}\dl{Fig. \ref{fig:roots} shows solutions $C$ for a linear profile} divided artificially into 4 layers \rev{($N=4$). Fig. \ref{fig:roots} shows the $N+1$ solutions for a given value of $kh$ for wave propagation along the axis of the direction of $U(z)$ ($\theta = 0$). 
For comparison, we show the two phase velocity solutions from the single layer ($N=1$) case as solid red lines. The phase velocity solutions from the $N=1$ case correspond} \dl{with equal vorticity in each layer, compared with the two known analytical solutions corresponding} to phase \rev{velocities} of plane waves propagating in directions $\mathbf{k}$ and $-\mathbf{k}$, respectively. \rev{For $N=4$ there are two solutions that agree with the \revs{well known exact solutions}, and three additional solutions that have phase velocities approximately equal the value of the current profile at the three layer interfaces, $U(-h/4)$, $U(-h/2)$, and $U(-3h/4)$ respectively. The vertical velocity profiles $w(z)$ for each of the phase velocity solutions is plotted in the small sub-figures in Fig. \ref{fig:roots}. For two velocity profiles corresponding to the solutions from the $N=1$, $w(z)$ is peaked at the surface, while for the three \revs{extra solutions}, $w(z)$ is peaked at layer interfaces $z = -h/4$, $z = -h/2$, and $z = -3h/4$. Given the constant vorticity of the fluid, the use of the PLA with $N>1$ \revs{cannot} change the physical nature of the problem, \revs{highlighting the spuriousness (in this context) of} the extra solutions.} 
\dl{For sufficiently long wavelengths (small $k$) the wave propagation speed exceeds the flow velocities in magnitude in both directions, and the physical solutions are simply that representing the highest velocity of each sign. For slower waves (higher $k$), however, there exists a level in the water column where, for either $\bk$ or $-\bk$, the flow velocity equals the phase velocity measured along the $x$ axis (the condition for a critical layer). This may occur when $\sqrt{(g/k)\tanh kh}<\Delta U$, where $\Delta U \equiv (\cos\theta)|\mathrm{max}[U(z)]-\mathrm{min}[U(z)]|$. 
In this case a technique is required for discarding the extra unphysical solutions.} 


\rev{\section{Nature of additional solutions: vorticity waves}

\label{sec:extrasol}
\begin{figure}[tb]
  \begin{center}
	  \includegraphics[width = 4in]{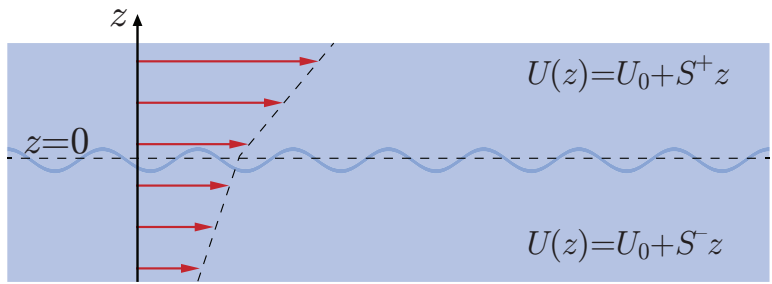}
  \end{center}
  \caption{Simplest model of a kink in the piecewise linear approximation. A sharp change in vorticity supports vorticity waves (or Rayleigh waves), not present for a smooth profile.}
  \label{fig:kink}
\end{figure}

The presence of sharp changes in vorticity allows the model system of Fig.~\ref{fig:geom} to support $N-1$ wave solutions of a different physical nature than the gravity waves at the free surface. These have been studied to some extent in the context of internal waves in the atmosphere\cite{holmboe62,caulfield94}, and are referred to as Rayleigh waves, from being first discussed by Lord Rayleigh\cite{rayleigh1879}, \revs{as} counter-propagating Rossby wave\revs{s}\cite{heifetz99} or a\revs{s} vorticity wave\revs{s}\cite{carpenter11}. A very readable review of the physical mechanism involved is found in section 4 of Ref.~\onlinecite{carpenter11}, and we shall only here recount a few main points in order to understand their appearence in the piecewise-linear model. 

Let us consider the simplest possible model of a kink in a doubly infinite piecewise linear velocity profile, as shown in Fig.~\ref{fig:kink}. Let the basic velocity be $U(z)=U_0 + S^\pm z$ so that $S^-$ and $S^+$ are the vorticities below and above the kink, respectively\revs{, and let the} fluid \revs{be} uniform. Assume moreover that the interface between the regions of different vorticity is slightly perturbed from $0$ to $\hat{\zeta}(x,t)\propto \exp(\rmi\bk\cdot\mathbf{r}-\rmi kCt)$. The Rayleigh equation again gives simple solutions for $w$ each side of the boundary: $w^\pm = A \rme^{\mp kz}$ 
with upper (lower) sign again denoting $z-\hat{\zeta}$ positive (negative). From the Euler equations we have,
similarly to Eq.~\eqref{eq:pressure}, 
\be
  k p^\pm/\rho = \rmi (C-U_0\cos\theta)(w^\pm)' + \rmi S^\pm \cos \theta w^\pm.
\ee
Demanding continuity of pressure at $z=0$ (linearized dynamic boundary condition) gives
\be\label{cdrift}
  C(\bk) = U_0\cos\theta + \frac{(S^+-S^-)\cos\theta}{2k}.
\ee
Noticing that $C= U_0\cos\theta$ would represent a perturbation that is simply drifting passively downstream, this wave mode has a nonzero intrinsic phase velocity when $S^+\neq S^-$. 

The vorticity wave is not a wave in the same sense as surface gravity waves, but is better thought of as a train of vortical structures which has come about due to perturbation $\hat{\zeta}$. Where $\hat{\zeta}<0$, fluid of vorticity $S^+$ is brought into the domain of background vorticity $S^-$, and \emph{vice versa}. The resulting train of vortical structures is instructively illustrated e.g.\ in figure 2 of Ref.~\onlinecite{heifetz99} and figure 4 of Ref.~\onlinecite{carpenter11}. (The vorticity equation also has an additional term when $\theta\neq 0,\pi$, due to undulations of the vortex lines of the background flow, see Ref.~\onlinecite{EllEJMB}).

The key point to notice in the present context is that such a wave mode can only be supported when there is a sharp change on a vertical lengthscale of a wave amplitude or less. Consequently, in a linear wave theory where wave amplitudes are infinitesimal, a smooth velocity profile will not support these modes. They are, in the present context, purely an artifact of the piecewise linear model. Note that although `spurious' in the system we consider here, vorticity waves \emph{can} be observed in other systems, such as atmospheric waves.

Given the artificial nature of the $N-1$ extra solutions in the context of linear waves considered herein, a technique for identifying and discarding them is necessary. The most natural method is the consider the resulting vertical velocity profiles shown in Fig. \ref{fig:roots}. The physical solutions corresponding to surface wave propagation have vertical velocity profiles peaked at the free surface, whereas in the case of the extra solutions, the velocity is peaked at a layer interface. By comparing $w(z)$ evaluated at the surface and interface heights for each of the $N+1$ solutions to the dispersion relation for a given $\mathbf{k}$, the desired surface wave solutions can be selected.

A simpler\revs{, pragmatic} method without considering $w(z)$ can be used for 
\revs{fast-moving wave modes which often occur for sufficiently} 
small values of \revs{$|\mathbf{k}|$, by} exploiting the property that extra solutions have phase velocities equal to the background flow at some depth\revs{, i.e., $\min[U(z)\cos\theta]\leq C\leq \max[U(z)\cos\theta]$}. 
For \revs{a range of} long wavelengths \revs{one or both of the desired phase velocities then exceed this range}, and \revs{can be immediately recognized as physical}. \revs{For monotonous $U(z)$ this will always be true for the shear-inhibited solution propagating along the current (It is possible in principle to construct a $U^\text{PL}(z)$ with very sharp kinks whose extra phase velocity solutions lie outside the range of $U(z)$. In keeping with the pragmatic nature of this method we may safely neglect this possibility since it does not occur for even very rough models of realistic flows.)}}

\section{Results}
The aim of this section is twofold. First, we verify and validate the numerical scheme as well as investigate the accuracy as a function of the number of layers. Secondly, we demonstrate the utility of the model in finding the dispersion relation for oft-occurring general profiles, and apply it to the ship wave problem, highlighting as an example of the use for Fourier transformation in the horizontal plane. We highlight some of the notable wave propagation characteristics that occur due to the curvature of the velocity profile as compared to a couette flow model.

\subsection{Verification and Validation}

\begin{figure}[ht]
  \includegraphics[scale=1.0]{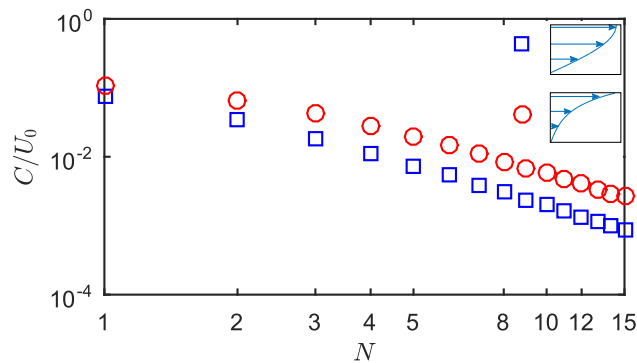}
  \caption{The phase velocity $C$ as a function of the number of layers $N$ evaluated at the wavevector $k_0$ satisfying the analytical dispersion relation (Eq.~\eqref{eq:perdr}) for stationary waves. Two current profiles of the form of (Eq.~\eqref{eq:uper}) were used with parameters $U_0/\sqrt{gh} = 0.45$,  $U_0^\prime = 0$ \rev{$\alpha=-0.62$, }(squares), and $U_0\sqrt{gh} = 0.45$,  $U_0^\prime/\sqrt{g/h} = $\dl{0.68}$\rev{1.36}$, \rev{$\alpha=2.23$, }(circles).}
\label{fig:conv}
\end{figure}

The convergence of the piecewise linear approximation 
when the number of layers increases 
has been proven 
in general 
by Zhang\cite{Zhang05}, and we 
verify it for our implementation as well.
We apply the piecewise linear approximation to a class of profiles where an analytical solution to the dispersion relation can be found for the 
special
case $C(\mathbf{k})=0$, analyzed by Peregrine\cite{Pere}:
\begin{equation}
U(z) = U_0\cosh\alpha^{1/2}z + U^\prime_0\alpha^{-1/2}\sinh\alpha^{1/2}z,
\label{eq:uper}
\end{equation} 
where $U_0$ and $U^\prime_0$ are the velocity and shear values at the surface respectively, and $\alpha$ is chosen here such that $U(-h) = 0$ for bottom depth $h$. Following Peregrine\cite{Pere} the 
wave number
$k_0$ satisfying 
$C(\mathbf{k}_0)=0$ 
solves 
the following equation:
\begin{equation}
(k_0^2 +\alpha)^{1/2}h\coth[(k_0^2 +\alpha)^{1/2}h] = (gh/U_0^2) + (U_0^\prime h/U_0).
\label{eq:perdr}
\end{equation}
To compare the N-layer model to this result, the phase velocity was evaluated 
by first choosing streamwise wave number
$k_x = -k_0$ found numerically from Eq.~\eqref{eq:perdr}, with the wavevector orientation opposite that of the current ($\theta = \pi$). The exact result for a smooth profile is $C = 0$ and the approximation error is shown in Fig.\ \ref{fig:conv} for \rev{a concave up profile with parameters $U_0/\sqrt{gh} = 0.45$,  $U_0^\prime = 0$, $\alpha=-0.62$, and a concave down profile with $U_0\sqrt{gh} = 0.45$,  $U_0^\prime/\sqrt{g/h} = 1.36$, and $\alpha=2.23$.} \dl{two different current profiles.} In both these cases, the phase velocity tends to zero $\sim N^{-2}$ in the limit of large $N$ in agreement with the result of Zhang\cite{Zhang05}.

When naively dividing the entire liquid column into equal layers, 
more layers are required to achieve a given level of accuracy for short wavelengths (large $kh$). 
This can be easily understood physically by noting that the influence on a regular wave from currents beneath the surface decreases expo\rev{n}entially with depth, and at the 1\% level dispersion properties are not influenced by currents deeper than a depth of $\lambda/2$, $\lambda$ being the wavelength. More quantitatively one could consider
the equation for the first-order correction to the phase velocity presented by 
Stewart and Joy%
\cite{StewartJoy} in infinite depth:
\begin{equation}
C \approx \sqrt{g/k} + 2k\int_{-\infty}^0U(z) e^{2kz} dz.
\label{eq:sj}
\end{equation}
Eq.~\eqref{eq:sj} is a weighted average of the current profile, with exponentially decreasing weight in the vertical direction. For shorter wavelengths, the layers in the PLA are more coarsely spaced in the depth-range where the weighting term is large, resulting in greater approximation error. An improvement in accuracy  is immediately achieved by diving the fluid into $N$ layers down to some intermediate depth determined by the desired accuracy. 
In our example, we divide the column into $N$ layers of equal width down to a depth equal to the smaller of $h$ and $\lambda/2$. When the column is cut off at $z=-\lambda/2$ this limits the accuracy that can be achieved: Since the weighting function $\exp(2kz)$ in the integral of Eq.~\eqref{eq:sj} is approximately $0.002$ at depth $\lambda/2$, even deeper waters must be included for accuracies better than the $10^{-3}$ level. The procedure works well, however, since this level of accuracy is achieved with only a small number of layers, typically $4$ or $5$ at the $1\%$ level. We thus achieve a computationally cheap scheme solving the dispersion problem with a uniform level of accuracy across the wave vector plane, with little variation in computational effort. 

\subsection{Dispersion relation}

\newcommand{\Cg}{\mathbf{C}_g}

\begin{figure*}[ht]
  \includegraphics[scale=1.0]{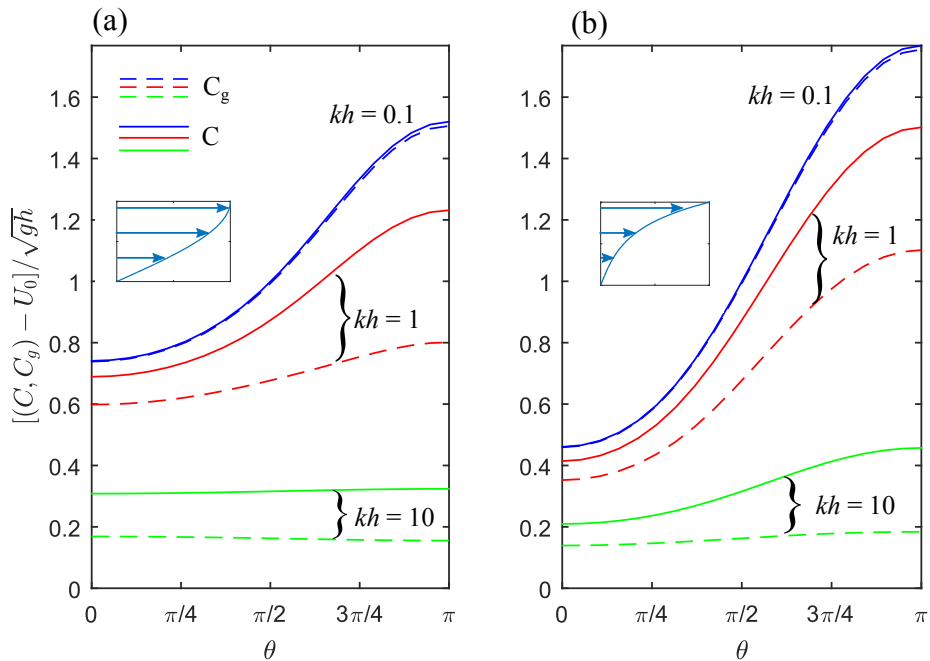}
  \caption{A comparison of the group and phase velocities relative to the surface velocity as a function of orientation angle relative to the current for two different current profiles. The current profile from (Eq.~\eqref{eq:uper}) was used with parameters $Fr_\text{sh} \equiv U_0/\sqrt{gh} = 0.45$, and $U_0^\prime = 0$ and $3U_0/h$ for the two profiles respectively.}
\label{fig:cg} 
\end{figure*}

In this section we compare the dispersion relation among two different profiles defined by Eq.~\eqref{eq:uper}. In particular we examine the phase velocity and group velocity, $
\Cg
= \nabla_\mathbf{k}\omega(\mathbf{k})$, as a function of propagation angle $\theta$ relative to the current direction, where 
$\nabla_\mathbf{k} = (\partial/\partial k_x, \partial/\partial k_y)$. 
In general the direction of 
$\Cg$ is not the same as the wavevector and phase velocity, and to simplify the analysis we consider here a scalar group velocity along the direction of $\mathbf{k}$, 
$C_g = \Cg\cdot\mathbf{k}/k$.  
For a more direct comparison, we define a shear Froude number 
$Fr_\text{sh} \equiv U_0/\sqrt{gh} = 0.45$, 
which is the same for both profiles. The concave-up profile then is prescribed $U_0^\prime = 0$, while the concave-down profile 
$U_0^\prime = 3U_0/h$. Three different wavelength scales relative to the depth are shown in Fig.\ \ref{fig:cg} corresponding to shallow, intermediate, and deep-water regimes. The phase velocities $C$ and group velocities $C_g$ relative to the surface velocity display different characteristics in each case. In shallow water ($kh = 0.1$), $C_g \approx C$ as usual since the medium becomes approximately non-dispersive, yet there is a directionally ($\theta$)-dependent propagation velocity magnitude due to the current profile. For the intermediate case ($kh = 1$), the latter remains true (to a lesser extent) but there is now a difference between the velocities the magnitude of which is also directionally dependent. In the deep water regime ($kh = 10$) the wavelength becomes small and the dispersion relation is influenced only by the near-surface current profile. For the concave up profile with zero surface shear, the well-known limit $C_g \approx C/2$ for deep water waves without current is approached, with the velocities being independent of propagation direction. This occurs to a lesser extent with the profile to the right due to finite surface shear.

\subsection{Example application: ship waves}
\rev{The effects of a background shear flow of constant vorticity on the behavior of ship wakes has been studied recently\citep{Ell14,Li16}, yet many shear profiles encountered in reality have depth-variable vorticity. A model taking into account arbitrary vorticity depth-dependence may be necessary in obtaining quantitatively accurate results for ship wakes and related parameters such as wave resistance in the presence of realistic shear flows. In this section we derive the solution to the ship wave problem with a piecewise linear background flow, and analyze the qualitative features of ship wakes in the presence of two illustrative example profiles of depth-varying vorticity approximately representative of flows encountered in reality.} 

The ship wave problem can be solved numerically for arbitrary shear profiles using the piecewise linear approximation in a direct generalization of the recent theory for the constant vorticity profile\cite{Ell14,Li16}. Assuming a stationary wave solution in a reference frame moving with the ship, a coordinate transformation ${\boldsymbol \xi} = \mathbf{x}-\mathbf{V}t$ is introduced where $\mathbf{V}$ is the velocity of a moving prescribed pressure source 
$\hat{p}_\text{ext}({\boldsymbol \xi})$ 
representing the ship relative to the undisturbed free surface. With this coordinate transformation used in the Fourier formulation, the time derivative becomes 
$\partial/\partial t = -i\mathbf{k}\cdot\mathbf{V}$. 
Thus, $-i\omega \rightarrow -i\mathbf{k}\cdot\mathbf{V}$ in the Euler equations (Eqs.~\eqref{eq:eulerx}-\eqref{eq:eulerz}). The reader is referred to
Refs.\ \onlinecite{Ell14,Li16} 
for more detail. 

The problem can be formulated in essentially the same way as described in section \ref{sec:der} by formulating a system of $N+1$ equations from matching conditions and boundary conditions. The velocity and pressure conditions are the same as Eq.~\eqref{eq:vmatch} and Eq.~\eqref{eq:pmatcha}-\eqref{eq:pmatchb}, respectively, save for replacing $-i\omega$ with $-i\mathbf{k}\cdot\mathbf{V}$ as explained above. A radiation condition $\mathbf{k}\cdot\mathbf{V} \rightarrow \mathbf{k}\cdot\mathbf{V} + i\epsilon$, $\epsilon\rightarrow 0+$ is necessary in practice to avoid singularities in the integral over the $\mathbf{k}$-plane.

The dynamic boundary condition is different from section \ref{sec:der}, where in the case of ship waves the pressure in Fourier space equals the prescribed pressure distribution $p_\text{ext}(\mathbf{k})$ (the Fourier transform of $\hat{p}_\text{ext}({\boldsymbol \xi})$) at the free surface. This leads to a non-zero term $p_\text{ext}(\mathbf{k})/\rho$ on the right side of Eq.~\eqref{eq:dkc}, which becomes:
\begin{align}
A_1\left[\gamma_1\sigma_\text{ship}\tanh kh_1 + \sigma_\text{ship}^2k + gk^2\tanh kh_1\right] +& \nonumber\\ 
B_1\left[\gamma_1\sigma_\text{ship} + \sigma_\text{ship}^2k\tanh kh_1 + gk^2\right] &= -\frac{k^2p_{ext}(\mathbf{k})}{\rho}\sigma_\text{ship},
\label{eq:dkcship}
\end{align}
where $\sigma_\text{ship} \equiv -i\mathbf{k}\cdot\mathbf{V} + ik_xU_0$.

The resulting $N+1$ equation system is inhomogeneous in this case and a solution to the coefficients $A_j$ and $B_j$ determining the vertical velocity can be found directly. The free surface in Fourier space representing the ship wake component at the given value of $\mathbf{k}$ is then found from the kinematic boundary condition:
\begin{equation}
\left(-i\mathbf{k}\cdot\mathbf{V} + ik_xU_0\right)\zeta = A_1\sinh kh_1 + B_1\cosh kh_1.
\label{eq:kbc}
\end{equation}
The real space pattern $\hat{\zeta}({\boldsymbol \xi})$ is found using an inverse fast Fourier transform.

\begin{figure*}[ht]
  \includegraphics[scale = 1.0]{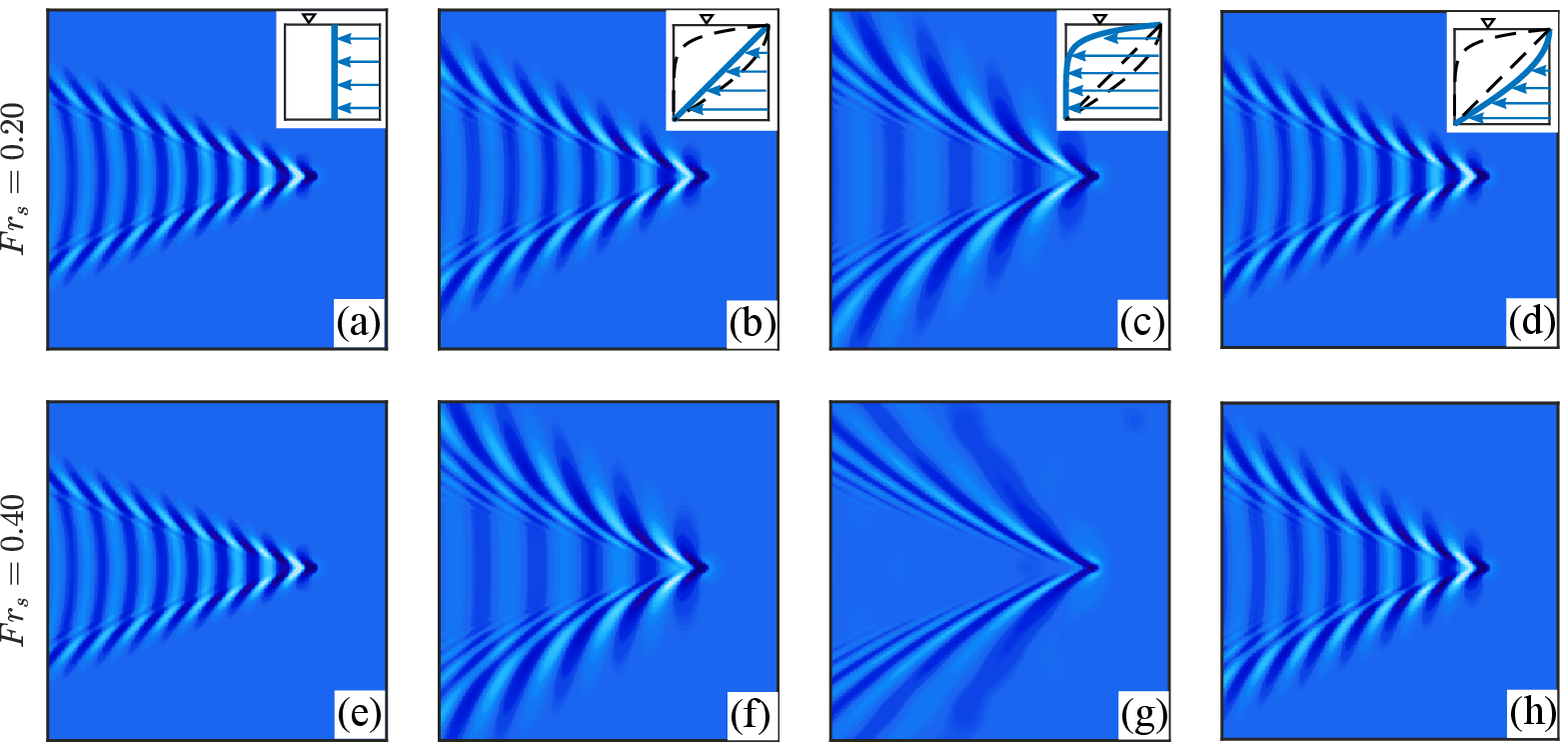}
  \caption{Ship wakes ($Fr = 0.6$, defined relative to the surface current) for different current profiles (columns) and shear Froude numbers $Fr_s$ (rows) for ship motion along the positive horizontal axis. A depth-uniform profile is shown in (a),\dl{(d)}\rev{(e)}, a linear profile of constant shear in (b),(f), \dl{and the piecewise linear approximation to the concave down current profile in (c),(f).}\rev{the exponential profile in (c),(g), and the concave up profile in (d),(h).} $Fr_s$ applies only to the \dl{two}\rev{three} rightmost columns, for which the \dl{surface}\rev{depth-averaged} shear value is the same. In all cases, depth $h$ is equal to the source radius $b$.}
\label{fig:wake}
\end{figure*}

To demonstrate the utility of the piecewise linear method for ship waves on an arbitrary current, we consider \rev{two current profiles of depth-varying vorticity. The first is} an exponential current profile defined as $U(z) = U_0\left(\exp[z/d]-1\right)$ with water depth $h = 10d$\rev{.}\dl{, approximated with $N=4$ layers.} \rev{The profile is similar to realistic wind-driven or river plume profiles with vorticity peaked at the surface and decaying with depth. The second profile is concave up defined by Eq.~\eqref{eq:uper} with equal $U_0$ and $h$ to the exponential profile, $U_0^\prime=0$, and $\alpha = -0.62$. Both profiles are approximated with $N=4$ layers.} To compare with the constant shear results in the literature, we here assign a shear Froude number $Fr_s \equiv VU_0/gh$, corresponding to the ``shear Froude number'' used in 
Refs.\ \onlinecite{Ell14,Li16}. 
In this sense we are assuming that the \dl{non-linear profile is}\rev{depth-variable vorticity profiles are} being compared with a constant model with vorticity equal to the depth-averaged value. We assume a Gaussian pressure distribution with half-width equal to water depth $b = h$ as $p_\text{ext}(\mathbf{\xi}) = p_0e^{-\pi^2\xi^2/b^2}$. 

The results of this comparison for Froude number $Fr = V/\sqrt{bg} = 0.6$ are shown in Fig.\ \ref{fig:wake}, along with the solution on a quiescent profile. Ship velocity $\mathbf{V}$ is expressed relative to the surface current and is assumed to be along the x-axis (as is the background flow). There are clear qualitative differences between the solutions. Insight can be gained by considering the transverse wavelengths, pertaining to the waves propagating parallel to the direction of ship motion in the central wake region. The transverse waves (having the same phase velocity as the source) in the \rev{three} rightmost columns are inhibited by the shear and therefore must be longer in wavelength in order to travel at the same speed as the corresponding waves in quescient waters. Furthermore, given the non-constant vorticity of the \rev{profiles in the two rightmost columns, the transverse wavelengths are notably changed from the constant vorticity case. For the concave down} exponential current profile, the current strength into the fluid is stronger relative to the constant shear approximation, which further increases the transverse wavelengths. \rev{For the concave up profile in the rightmost column, the vorticity is zero at the surface and the current strength weaker than the constant vorticity model, resulting in shorter transverse wavelengths.}  The differences are exaggerated in the bottom row, where the shear is strong enough such that $Fr$ is supercritical \rev{for the exponential profile in (g)}, disallowing transverse waves (see 
Ref.\ \onlinecite{Li16} 
for details). For \dl{the current profile}\rev{the current profiles} considered here, inclusion of the depth-varying vorticity of the profile is essential to obtain a realistic solution to the wake, and consequently the wave resistance, of the ship.

\section{Conclusion}

This study has demonstrated the use of a piecewise linear approximation
(PLA)
to an arbitrary current profile to model \rev{linear} wave propagation in 3D. The method is valid for any current magnitude and wavelength and does not rely on assumptions of weak current, near-potentiality, or small vorticity as do many other approximation techniques presented in the literature. The approximation accuracy is relatively unchanged over 
all
wavelengths making the technique well suited for 
solving problems formulated in 
Fourier
wavevector space, where integration over the full 
plane of wave vectors
is necessary. \rev{The accuracy of the PLA method could be further verified through comparison to experimental data of the dispersion relation where independent measurements of the background flow are performed, such as in Lund \textit{et al}.\cite{Lund15}.} Additional \dls{analysis}\revs{discussion} concerning the nature and characteristics of extra spurious solutions to the PLA is given showing that the\dls{spurious}\dls{$N-1$ extra} solutions \revs{to the dispersion relation which the PLA produces}
represent vortical structures flowing along at a velocity near that of the layer interface. \revs{They are artifacts of the discontinuities in vorticity introduced by the model, are spurious in the present context, and should be discarded. Details of the procedures used for quickly identifying the two physical solutions for a given $\bk$ are provided.}

The importance of including non-uniform vorticity into wave propagation models is highlighted by considering the directional-dependence of the group and phase velocities
for concave-up vs concave-down profiles,
as well as solution to the ship wave problem.

The PLA in 3D is a practical method for solving a wide variety classical wave problems such as ship waves as elaborated upon herein, ring waves, and problems of radiation and refraction,
in the presence of a shear current of arbitrary depth-dependence.
Further extensions of the model could include velocity profiles where the direction varies with depth, as well as non-linear waves as been done by e.g.\ Dalrymple\cite{Dalrymple74} and Swan \textit{et al.}\cite{Swan01}.

\begin{acknowledgments}
The quadratic eigenvalue formulation of our PLA algorithm described briefly in section \ref{sec:der} was conceived and implemented by Peter Maxwell. We gratefully acknowledge his efforts, expecting to draw on this improved capability in several future applications. S\AA E was partly funded by the Norwegian Research Council (FRINATEK), project 249740. BKS is funded by the Department of Energy and Process Engineering, NTNU.
\end{acknowledgments}

\bibliography{nlayer_references}

\end{document}